\documentclass[aps,pra,twocolumn,showpacs,superscriptaddress,10pt]{revtex4-1}

\usepackage{xcolor, graphicx}
\usepackage{amsmath, amssymb}
\usepackage{ulem}
\usepackage[colorlinks=true,urlcolor=blue,citecolor=blue,linkcolor=blue]{hyperref}

\begin{document}

\title{Characterizing maximally singular phase-space distributions}

\author{J. Sperling}\email{jan.sperling@uni-rostock.de}
\affiliation{Arbeitsgruppe Theoretische Quantenoptik, Institut f\"ur Physik, Universit\"at Rostock, D-18051 Rostock, Germany}

\begin{abstract}
	Phase-space distributions are widely applied in quantum optics to access the nonclassical features of radiations fields.
	In particular, the inability to interpret the Glauber-Sudarshan distribution in terms of a classical probability density is the fundamental benchmark for quantum light.
	However, this phase-space distribution cannot be directly reconstructed for arbitrary states, because of its singular behavior.
	In this work, we perform a characterization of the Glauber-Sudarshan representation in terms of distribution theory.
	We address important features of such distributions:
	(i) the maximal degree of their singularities is studied,
	(ii) the ambiguity of representation is shown,
	and (iii) their dual space for nonclassicality tests is specified.
	In this view, we reconsider the methods for regularizing the Glauber-Sudarshan distribution for verifying its nonclassicality.
	This treatment is supported with comprehensive examples and counterexamples.
\end{abstract}

\date{\today}
\maketitle

\section{Introduction}
	Since the theoretical explanation of the photoelectric effect~\cite{E05}, quantum optics has developed into a main field of modern physics.
	A major focus is on the theoretical description of quantized radiation fields and the characterization of genuine quantum effects~\cite{VW06}.
	Nowadays, quantum light is applied, for example, to perform secure communication protocols~\cite{GT07}.

	A fundamental feature of the quantum optical description of light is the representation in phase space.
	Employing such a representation, a single-mode light field can be characterized and evolves formally in the same same way as a classical harmonic oscillator.
	Such a semi-classical approach connects a quantum state, given by a density operator $\hat \rho$, with a corresponding phase-space distribution.
	Prominent examples are the Wigner distribution~\cite{W32}, the Husimi distribution~\cite{H40}, and the Glauber-Sudarshan (GS) or $P$ distribution~\cite{S63,G63}, which yields a diagonal expansion in terms of coherent states $|\alpha\rangle$:
	\begin{align}\label{eq:GlauberSudarshan}
		\hat\rho=\int d^2\alpha\, P(\alpha)|\alpha\rangle\langle\alpha|.
	\end{align}

	The GS distribution even defines the notion of nonclassicality~\cite{TG65,M86}.
	Namely, for nonclassical states, it cannot be interpreted in terms of a classical probability, which refers to as a quasiprobability distribution.
	Such distributions can have negative contributions in two forms.
	If the GS distribution is regular, i.e., a continuous function, the negativities can be directly observed: $P(\alpha)<0$ for some $\alpha$.
	If the $P$ distribution is singular, we have negativities in the sense of distributions, which will be studied in this work.

	In general, there exist a number of techniques to reconstruct phase-space distributions; see~\cite{WVO99,LR09} for reviews.
	However, the GS function is, in general, not accessible because of its singularities.
	Therefore, some regularization methods have been proposed~\cite{K66,AW70}.
	Among others, the $s$-parametrized quasiprobabilities~\cite{CG69} have been mostly reconstructed in experiments.
	However, due to convolution, some of the nonclassical effects cannot be detected with such phase-space functions; see, e.g.,~\cite{K13}.
	For example, the Wigner function ($s=0$) has the properties of a classical probability density for squeezed states and the Husimi function ($s=-1$) is a classical probability distribution for any state.
	To overcome this deficiency, non-Gaussian filter methods have been proposed~\cite{AW70}, which regularize the $P$~distribution for any state and exhibit negativities for any nonclassical state~\cite{KV10}.
	Applying this approach in experiments, one can even directly uncover the nonclassicality of squeezed states~\cite{KVHS11,ASVKMH15}.

	In contrast to the above attempts to regularize the singularities of the GS distribution, one can alternatively study its Fourier transform.
	This so-called characteristic function is regular for any state and it can be directly sampled in experiments~\cite{LS02,ZPB07,MKMPE11}.
	Moreover, a hierarchy of necessary and sufficient nonclassicality conditions has been proposed on this basis~\cite{V00,RV02}.
	Similarly to nonclassicality tests in terms of moments, e.g., in~\cite{SV05,MBWLN10}, the characteristic function allows one a full characterization of the nonclassicality of a quantum state of light.
	Recently, the moment-based and characteristic function nonclassicality conditions were even unified~\cite{RSAMKHV15}.

	The GS distribution has been further generalized in Ref.~\cite{V08} to capture time-dependent quantum effects.
	Moreover, it inspired the introduction of optimized entanglement quasiprobabilities that are negative if and only if the a state is entangled~\cite{STV98,SV09}.
	In this context, it is also worth mentioning that an entangled state of light requires a nonclassical GS distribution.
	Additionally, the regularization of single-mode quasiprobabilities has been extended to multipartite systems to verify quantum correlations beyond entanglement~\cite{FP12,ASV13}.

	In this paper, the singularities of the GS distribution are characterized.
	Applying the technique of Sudarshan, it is demonstrated that the representation of the $P$ distribution is not unique.
	Using the characteristic function, the singularities of the GS distribution are bounded and a maximally singular GS distribution is established.
	This exponential function of second-order derivatives of the Dirac $\delta$ distribution is shown to present, in some sense, a strict bound to the irregularities for all quantum states of light.
	Based on this finding, the regularization and the dual space of the GS distributions are considered.
	A number of examples are given.
	Thus, these rigorous and analytical studies give a deeper insight into the GS quasiprobability and the related characterization of nonclassical states in optical systems.

	The paper is structured as follows.
	In Sec.~\ref{sec:GSDistributions}, we expand $P$ distributions in terms of singular distributions and identify the maximally singular distribution in this class.
	An extended discussion and examples are given in Sec.~\ref{sec:DnE}.
	Section~\ref{sec:FilteredP} visualizes different singular phase-space distributions in terms of regular functions.
	Nonclassicality criteria and the underlying test functions are considered in Sec.~\ref{sec:Dual}.
	A summary and conclusions are given in Sec.~\ref{sec:SnC}.

\section{Glauber-Sudarshan distributions}\label{sec:GSDistributions}
	The question of nonclassicality has been completely solved for pure states~\cite{C65,H86}.
	The classical and pure states are exclusively coherent ones~\cite{S26},
	\begin{align}\label{eq:CoherentState}
		|\alpha\rangle=\hat D(\alpha)|{\rm vac}\rangle=e^{-|\alpha|^2/2}e^{\alpha\hat a^\dagger}|{\rm vac}\rangle.
	\end{align}
	Here, $\hat D(\alpha)=\exp(\alpha\hat a^\dag-\alpha^\ast\hat a)$ is the displacement operator, $\hat a$ ($\hat a^\dag$) is the annihilation (creation) operator, and $|{\rm vac}\rangle$ denotes the vacuum state.

	For mixed states, the question of weather or not a system is classical is a more involved problem, due to the singularities of the GS distribution.
	An example of a nonclassical and singular GS distribution is
	\begin{align}
		P(\alpha)=\left(1+\eta\partial_\alpha\partial_{\alpha^\ast}\right)\delta(\alpha),
	\end{align}
	which describes a single-photon state that is mixed ($0<\eta\leq1$) with vacuum~\cite{LS02}.
	This $P$ distribution is clearly highly singular due to the second-order derivative of the Dirac $\delta$~distribution.
	For regular (smooth) GS functions, the nonclassicality can be directly visualized.
	For instance, the following $P$ distribution is negative for $\alpha=0$ (for any $\bar n>0$):
	\begin{align}
		P(\alpha)=\frac{1}{\pi \bar n^3}\left[(\bar n+1)|\alpha|^2-\bar n\right]\exp\left(-\frac{|\alpha|^2}{\bar n}\right).
	\end{align}
	It represents a so-called single-photon-added thermal state~\cite{AT92}, i.e., a single photon on a thermal background, whose phase-space $P$ distribution has been experimentally reconstructed~\cite{KVPZB08}.
	These examples justify that a rigorous analysis of the singularities of the GS distributions is indispensable for a profound understanding of the nonclassical features of mixed quantum states of light.

\subsection{Quantum state representations}
	Apart from those particular examples and to characterize the distribution of general quantum states, let us study the Fourier transform of the GS distribution
	\begin{align}\label{eq:CF}
		\Phi(\beta)=\langle{:}\hat D(\beta){:}\rangle=\int d^2\alpha\, P(\alpha) e^{\beta\alpha^\ast-\beta^\ast\alpha},
	\end{align}
	which is referred to as characteristic function and where the normally ordered displacement operator reads ${:}\hat D(\beta){:}=\exp(\beta\hat a^\dag)\exp(-\beta\hat a)$.
	The inverse Fourier transform is
	\begin{align}
		P(\alpha)=\frac{1}{\pi^2}\int d^2\beta\, \Phi(\beta) e^{\beta^\ast\alpha-\beta\alpha^\ast}.
	\end{align}
	Any quantum state $\hat\rho$ may be also described by its density operator in the Fock basis
	\begin{align}\label{eq:DensityOp}
		\hat\rho=\sum_{m,n=0}^\infty \rho_{m,n} |m\rangle\langle n|.
	\end{align}
	In the first step, the distributions that describe the basis elements $|m\rangle\langle n|$ will be considered.
	Rigorously speaking, we adopt the approach of Sudarshan~\cite{S63}, who showed that for any quantum state the representation~\eqref{eq:GlauberSudarshan} exists, in Fourier space.
	Note that this work is restricted to single-mode considerations only.
	However, a generalization to multimode scenarios is straightforward.

	The characteristic function~\eqref{eq:CF} of $|m\rangle\langle n|$ reads
	\begin{align}\label{eq:PFockCF}
	\begin{aligned}
		\Phi_{m,n}(\beta)=&\langle n|{:}\hat D(\beta){:}|m\rangle
		\\=&\sum_{k=0}^{\min\{m,n\}}\frac{
			\sqrt{m!n!}\,
			\beta^{n-k}
			(-\beta^\ast)^{m-k}
		}{k!(m-k)!(n-k)}.
	\end{aligned}
	\end{align}
	Applying the inverse transformation, one gets the corresponding $P$ distribution as
	\begin{align}\label{eq:PFock}
		P_{m,n}(\alpha)=\sum_{k=0}^{\min\{m,n\}}\frac{\sqrt{m!n!}(-1)^{m+n}}{k!(m-k)!(n-k)!}
		\partial_{\alpha}^{m-k}\partial_{\alpha^\ast}^{n-k}\delta(\alpha).
	\end{align}
	Thus, we can formally write for any Fock basis element $|m\rangle\langle n|=\int d^2\alpha\, P_{m,n}(\alpha) |\alpha\rangle\langle \alpha|$.
	Using the Fock expansion~\eqref{eq:DensityOp} of $\hat \rho$, we obtain the $P$ function in the form
	\begin{align}\label{eq:Pexpansion}
		P(\alpha)=&\sum_{m,n=0}^\infty \rho_{m,n}P_{m,n}(\alpha).
	\end{align}
	We will discuss this decomposition in detail later on.

\subsection{Maximally singular distribution}
	The unitary displacement and the normally ordered displacement operators are related via the Baker-Campbell-Hausdorff formula ${:}\hat D(\beta){:}=\exp(|\beta|^2/2)\hat D(\beta)$.
	As the modulus of any unitary operator is bounded by one, one finds with the above relation for the characteristic function~\cite{V00}
	\begin{align}\label{eq:MaxIncreaseCF}
		|\Phi(\beta)|\leq e^{-|\beta|^2/2}.
	\end{align}
	Let us also emphasize that it has been shown in~\cite{V00} that for all classical states $|\Phi(\beta)|\leq 1$ holds.
	Moreover, one should recall that the asymptotic behavior in Fourier space is closely related to the regularity in the original space.
	That is, in terms of Sobolev's lemma~\cite{Y80}, if $|\beta|^{2r}\Phi(\beta)$ is an integrable function, $\int d^2\beta\,|\beta|^{2r}|\Phi(\beta)|<\infty$, then $\partial_\alpha^r\partial_{\alpha^\ast}^r P(\alpha)$ is continuous.

	Conversely, the maximal increase in inequality~\eqref{eq:MaxIncreaseCF} bounds the maximal singularity of the GS distribution.
	As the phase $\arg \Phi(\beta)$ does not affect this bound, one may write the characteristic function of the worst-case distribution as
	\begin{align}
		\Phi_{\max}(\beta)=\exp\left(\frac{|\beta|^2}{2}\right).
	\end{align}
	Applying the inverse Fourier transform, the corresponding $P$ distribution is consequently
	\begin{align}\label{eq:Pmax}
		P_{\rm max}(\alpha)=\exp\left(-\frac{1}{2}\partial_\alpha\partial_{\alpha^\ast}\right)\delta(\alpha).
	\end{align}
	The discussion of this distribution will be done in the following sections.
	In the rest of this work, we will refer to a distribution that is at most as singular as $P_{\max}(\alpha)$ as a GS distribution.

\section{Discussions and Examples}\label{sec:DnE}
	Combining Eqs.~\eqref{eq:Pexpansion} and~\eqref{eq:PFock}, one directly observes that any GS distribution, which describes a quantum state, has the formal structure
	\begin{align}\label{eq:SingularGS}
		&P(\alpha)
		\\=&\sum_{k=0}^{\infty}\sum_{m,n=k}^\infty \rho_{m,n}\frac{\sqrt{m!n!}(-1)^{m+n}}{k!(m-k)!(n-k)!}
		\partial_{\alpha}^{m-k}\partial_{\alpha^\ast}^{n-k}\delta(\alpha).
		\nonumber
	\end{align}
	Here let us already mention that the sum yields, in general, an infinite order of derivatives of the Dirac $\delta$~distribution, which is substantially different from any finite order, as we will observe in detail later on.
	In this section, we will further characterize these types of distributions properly.
	Examples will underline the features of GS distributions.
	In addition, some representations of $P_{\max}$ are also considered.

\subsection{General comments}
	Let us study some properties of the elements $P_{m,n}(\alpha)$ in Eq.~\eqref{eq:PFock} and their combinations in Eq.~\eqref{eq:SingularGS}.
	The symmetry
	\begin{align}
		P_{m,n}(\alpha)=P_{n,m}(\alpha)^\ast
	\end{align}
	is a direct consequence of the relations $(\partial_{\alpha}^q\partial_{\alpha^\ast}^r)^\ast=\partial_{\alpha^\ast}^q\partial_{\alpha}^r$ and $\delta(\alpha)=\delta(\alpha)^\ast$.
	The normalization reads
	\begin{align}
		\int d^2\,\alpha P_{m,n}(\alpha)=\delta_{m,n}
	\end{align}
	by using the integral $\int d^2\alpha\, \partial_{\alpha}^q\partial_{\alpha^\ast}^r\delta(\alpha)=\delta_{q,0}\delta_{r,0}$ and with $\delta_{m,n}$ denoting the Kronecker symbol.
	Combining both results with the properties of the Fock matrix elements $\rho_{m,n}=\rho_{n,m}^\ast$ and $\sum_{n=0}^\infty \rho_{n,n}=1$, we find the well known facts that the $P$ distribution in the form~\eqref{eq:SingularGS} is real-valued (in the sense of distribution theory), $P(\alpha)=P(\alpha)^\ast$, and normalized, $\int d^2\alpha\, P(\alpha)=1$~\cite{S63}.

	Moreover, each of the distributions $P_{m,n}(\alpha)$ has the support of a single element.
	That is, $P_{m,n}(\alpha)=0$ for all $\alpha\neq0$.
	This can be directly observed by applying a sufficiently smooth test function $F(\alpha)$,
	\begin{align}\label{eq:finiteDual}
		&\int d^2\alpha\, F(\alpha)P_{m,n}(\alpha)
		\\=&\sum_{k=0}^{\min\{m,n\}}\frac{\sqrt{m!n!}(-1)^{m+n}}{k!(m-k)!(n-k)!}
		\left[\partial_{\alpha}^{m-k}\partial_{\alpha^\ast}^{n-k}F(\alpha)\right]_{\alpha=0}.
		\nonumber
	\end{align}
	This expression solely depends on the local properties of $F(\alpha)$ at the point $\alpha=0$.
	The general features of proper test functions are further studied in Sec.~\ref{sec:Dual}.

\subsection{Fock representation of $P_{\max}$}
	The Fock representation of an operator $\hat\mu$ that is given by the distribution in Eq.~\eqref{eq:Pmax} is
	\begin{align}
		\hat\mu=&\int d^2\alpha\, P_{\max}(\alpha)|\alpha\rangle\langle\alpha|
		\\=&\sum_{m=0}^\infty \frac{(-1/2)^m}{m!}\left[\partial_\alpha^m\partial_{\alpha^\ast}^m e^{-|\alpha|^2}e^{\alpha\hat a}|{\rm vac}\rangle\langle{\rm vac}|e^{\alpha^\ast\hat a}\right]_{\alpha=0}
		\nonumber
		\\=&\sum_{m=0}^\infty \left(-\frac{1}{2}\right)^m\sum_{k=0}^m\frac{m!}{k!(m-k)!} |k\rangle\langle k|
		\nonumber
	\end{align}
	where we applied the Leibniz rule (product rule for higher derivatives) and used $\hat a^{\dagger r}|{\rm vac}\rangle=\sqrt{r!}|r\rangle$.
	This expression for $\hat\mu$ can be further simplified,
	\begin{align}
		\hat\mu=&\sum_{k=0}^\infty \frac{(-1/2)^k}{k!}\left[\sum_{m=k}^\infty \frac{m!}{(m-k)!}\gamma^{m-k} \right]_{\gamma=-1/2} |k\rangle\langle k|
		\nonumber
		\\=&\sum_{k=0}^\infty \frac{(-1/2)^k}{k!}\left[\partial_\gamma^k\frac{1}{1-\gamma}\right]_{\gamma=-1/2} |k\rangle\langle k|
		\nonumber
		\\=&\sum_{k=0}^\infty \frac{2(-1)^k}{3^{k+1}} |k\rangle\langle k|,
	\end{align}
	with $\partial_\gamma^k(1-\gamma)^{-1}=k!(1-\gamma)^{-1-k}$.

	From this form of the operator $\hat\mu$, being the Fock representation to the distribution $P_{\max}$ with an infinite rank, we can see that $\hat\mu$ is neither normalized, ${\rm tr}(\hat\mu)=1/2$, nor a positive operator (negative eigenvalues for odd $k$ values).
	Therefore, it does not represent a valid quantum state.
	However, the operator
	\begin{align}\label{eq:AbsMu}
		|\hat\mu|=\sum_{k=0}^\infty \frac{2}{3^{k+1}} |k\rangle\langle k|
	\end{align}
	represents a valid quantum state, that is, a thermal state with a mean thermal photon number $\bar n=1/2$.
	It should be stressed that the thermal state, which is defined in Eq.~\eqref{eq:AbsMu}, will be used for comparing different distributions later on.

\subsection{$s$-Parametrized representation of $P_{\max}$}\label{subsec:sParametrization}
	The characteristic functions of $s$-parametrized quasiprobabilities $\Phi(\beta;s)$ and the one of the $P$ distribution ($s=1$) can be related~\cite{VW06},
	\begin{align}
		\Phi(\beta;s)=\exp\left(-\frac{1-s}{2}|\beta|^2\right)\Phi(\beta).
	\end{align}
	Thus, we have for the maximal GS distribution an $s$-parametrized characteristic function
	\begin{align}
		\Phi_{\max}(\beta;s)=\exp\left(\frac{s}{2}|\beta|^2\right),
	\end{align}
	and its Fourier transform
	\begin{align}\label{eq:sParametrisedPmax}
		P_{\max}(\alpha;s)=\exp\left(-\frac{s}{2}\partial_\alpha\partial_{\alpha^\ast}\right)\delta(\alpha).
	\end{align}
	See also Appendix~\ref{app:ExponentialLaplace} for further details on exponential operators of $\partial_\alpha\partial_{\alpha^\ast}$.
	In the form~\eqref{eq:sParametrisedPmax}, it can be seen for the Wigner representation ($s=0$) that the maximally singular Wigner distribution is a Dirac $\delta$~distribution
	\begin{align}\label{eq:maxWigner}
		W_{\max}(\alpha)=P_{\max}(\alpha;0)=\delta(\alpha). 
	\end{align}
	This supports the result that the Wigner function is regular for any quantum state, as $W_{\max}$ of the maximally singular distribution $P_{\max}$ has ``only'' the singularity of a Dirac $\delta$~distribution.

	For any $s<0$, it can be also observed that the Fourier transform is not unique for general GS distributions.
	Namely, we have $\Phi_{\max}(\beta;s)=\exp(-|s||\beta|^2/2)$ in this case, which also has the inverse Fourier transform:
	\begin{align}\label{eq:negsParametrisedPmax}
		P_{\max}(\alpha;s)=\frac{2}{\pi|s|}\exp\left(-\frac{2|\beta|^2}{|s|}\right)
		\quad\text{for }s<0.
	\end{align}
	In the following we discuss the relation between Eqs.~\eqref{eq:sParametrisedPmax} and~\eqref{eq:negsParametrisedPmax} on the basis of an example.

\subsection{A singular and regular example}
	We may consider a thermal state described by the density matrix $\rho_{m,n}=\delta_{m,n}(\bar n+1)^{-1}[\bar n/(\bar n+1)]^m$ for a mean thermal photon number $\bar n>0$, i.e.,
	\begin{align}\label{eq:FockThermal}
		\hat\rho=\frac{1}{\bar n+1}\sum_{m=0}^\infty \left(\frac{\bar n}{\bar n+1}\right)^m|m\rangle\langle m|.
	\end{align}
	Its characteristic function is $\Phi(\beta)=\exp(-\bar n|\beta|^2)$.
	Inserting the Fock expansion~\eqref{eq:FockThermal} into relation~\eqref{eq:SingularGS} yields
	\begin{align}\label{eq:ThermalSingular}
		P(\alpha)=&\exp\left(\bar n\partial_\alpha\partial_{\alpha^\ast}\right)\delta(\alpha).
	\end{align}
	In contrast, it was shown for thermal states that we have a regular Gaussian $P$ distribution~\cite{G63}
	\begin{align}\label{eq:ThermalRegular}
		P(\alpha)=&\frac{1}{\pi\bar n}\exp\left(-\frac{|\alpha|^2}{\bar n}\right),
	\end{align}
	similarly to the previous discussion of the $s$-parametrized $P$ representation for $s<0$.

	The mathematical interpretation of the ambiguity of Eqs.~\eqref{eq:ThermalSingular} and~\eqref{eq:ThermalRegular} is simple.
	The same distributions can be represented in different forms; see also~\cite{BV87}.
	However, in many papers one can read statements similar to the following: `A nonclassical state is described by a negative $P$ function or a distribution which is more singular than the Dirac $\delta$~distribution.'
	Considering the example of a classical thermal state at hand [Eq.~\eqref{eq:ThermalSingular}], one can see that such a statement is incorrect.
	Having a closer look at the original definition of nonclassicality, it is stated that classical states are those ``...fields which have a positive-definite $P(\alpha)$...'' distribution~\cite{TG65}.
	That is, independently of the representation, if a clearly non-negative $P$ representation exists [for example, in Eq.~\eqref{eq:ThermalRegular}], the state is classical.
	Moreover, the definition relates to the distribution as a linear map on test functions $f(\alpha)$, e.g., $\int d^2\alpha\, P(\alpha)|f(\alpha)|^2\geq0$ for classical $P(\alpha)$, which is independent of the representation of $P$ (see also Sec.~\ref{sec:Dual}).

	Finally, let us make another remark.
	The confusion with the singularities of the $P$ distribution in the literature is due to the fact the general decomposition~\eqref{eq:SingularGS} and those in Ref.~\cite{S63} require a sum of all orders of derivatives of the Dirac $\delta$~distribution, which are typically not taken into account.
	In the case of finite distributions $\rho_{m,n}=0$ for $m,n>n_{\max}$, we can apply the result in~\cite{V00} to see that, indeed, the corresponding states are necessarily nonclassical if finite orders of derivatives of $\delta(\alpha)$ are involved.
	As the characteristic function $\Phi(\beta)$ in such a finite case is a polynomial [cf.~\eqref{eq:PFock}], which is not constant, it will necessarily exceed the classical limit for some large $\beta$ value, i.e., $|\Phi(\beta)|>1$.

\subsection{A strict bound for physical $P$ distributions}
	Previously, we have shown that the operator $\hat\mu$ does not represent a physical quantum state.
	Here, however, it is shown that $P_{\max}$ can be considered as a bound to any quantum state.
	In particular, one can demonstrate that the upper bound in Eq.~\eqref{eq:MaxIncreaseCF} can be approached by physical states for any direction in phase space.

	For this aim, we consider, up to a rotation and a displacement in phase space, a general pure squeezed state
	\begin{align}\label{eq:SqStateFock}
		|\xi\rangle=\frac{1}{\sqrt{\cosh\xi}}\exp\left(-\frac{\tanh\xi}{2}\hat a^{\dagger 2}\right)|{\rm vac}\rangle,
	\end{align}
	which is characterized by the squeezing parameter $\xi>0$.
	Compared to the variance of the field fluctuations of the vacuum state, this squeezed state has a noise reduction (amplification) of a factor $e^{-2\xi}$ ($e^{2\xi}$) in the squeezed (antisqueezed) direction of phase space~\cite{VW06}.
	The characteristic function of this squeezed state reads
	\begin{align}\label{eq:SqueezedStateCF}
		\Phi(\beta)=\exp\left(-\sinh^2\xi|\beta|^2-\cosh\xi\sinh\xi\frac{\beta^2+\beta^{\ast2}}{2}\right).
	\end{align}
	Choosing the particular direction $\beta=i|\beta|$, we have
	\begin{align}\label{eq:SqueezedStateCFmax}
		\Phi(i|\beta|)=\exp\left(\left[1-e^{-2\xi}\right]\frac{1}{2}|\beta|^2\right).
	\end{align}

	Thus, for $\xi\to\infty$, we approach the rising behavior of $\Phi_{\max}(i|\beta|)=\exp(|\beta|^2/2)$.
	A rotation in phase space, $\hat a\mapsto\hat a e^{i\varphi}$, allows us to make the same statement for any direction.
	However, this cannot be done for all directions simultaneously, as we have for our specific example~\eqref{eq:SqueezedStateCF} in the direction $\beta=|\beta|$ a decaying characteristic function, which reads $\Phi(|\beta|)=\exp(-[e^{2\xi}-1]|\beta|^2/2)$.
	Finally, we can conclude from the Fourier transform of Eq.~\eqref{eq:SqueezedStateCF}, see also~\cite{KVHS11} or Appendix~\ref{app:ExponentialLaplace}, that any direction of $P_{\max}$ in phase space can be considered as the limiting distribution of an infinitively squeezed state in this phase-space direction.
	This also relates to our earlier observation that the Wigner function of the maximal GS distribution [Eq.~\eqref{eq:maxWigner}] is described by a Dirac $\delta$~distribution, which resembles a ``state'' with an infinite squeezing in all phase-space directions.

\subsection{Preliminary summary}
	In this section, the characterization of singular GS distributions has been performed.
	First, the state representation in terms of a sum over all derivatives of the Dirac $\delta$~distribution has been studied.
	We showed that such a representation is ambiguous in the sense that a regular and a highly singular distribution may describe the same state.
	This was discussed on the basis of the $s$-parametrized quasiprobabilities as well as the example of a thermal state.
	Second, the features of the most singular GS distributions $P_{\max}$ have been considered.
	In particular, the Fock expansion and the $s$ parametrization of such a distribution have been established.
	Based on the example of a squeezed state, we showed that for any direction in phase space, this maximally singular distribution can be approached.
	Thus, it represents a strict bound to the singularities of any $P$ representation.

\section{Regularized phase-space distributions}\label{sec:FilteredP}
	The approximation of a singular phase-space distribution with regular ones has always been a subject of many investigations.
	For instance, it has been shown that any $P$ distribution can be retrieved in the limit of smooth~\cite{K66} or square integrable~\cite{KMC65} functions.
	A general method to access regular phase-space functions was formulated in Ref.~\cite{AW70} and further specified in Ref.~\cite{KV10}.
	In the latter approach, the regularization of $P$ distributions is done such that it can unconditionally uncover the nonclassicality of light via regular, so-called, filtered quasiprobabilities.
	Moreover, it was shown that, for certain filter functions, these filtered quasiprobabilities are not only continuous, but also smooth functions~\cite{ASV13}.

	Here we aim at applying this filter approach to characterize the distribution $P_{\max}$.
	Therefore, we briefly recapitulate the approach in~\cite{KV10} with slightly different notation.
	The filtered quasiprobability is the convolution
	\begin{align}\label{eq:FilterPconv}
		P_{\Omega}(\alpha;w)=\int d^2\alpha_0\, P(\alpha-\alpha_0)\Omega(\alpha_0;w),
	\end{align}
	where $\Omega(\alpha;w)$ represents, for any parameter $w>0$, a sufficiently well-behaved positive-semi-definite probability density that has a limit $\lim_{w\to\infty}\Omega(\alpha;w)=\delta(\alpha)$.
	It was shown~\cite{KV10} that these requirements are fulfilled if $\Omega(\alpha;w)$ is the inverse Fourier transform of a function
	\begin{align}
		\tilde\Omega(\beta;w)=&\frac{\tilde\Omega'(\beta/w)}{\tilde\Omega'(0)},
		\text{ with}\\
		\tilde\Omega'(\beta)=&\int d^2\beta_0\, \omega(\beta+\beta_0)\omega(\beta)^\ast,
		\nonumber
	\end{align}
	with a function $\omega(\beta)$ that decays sufficiently fast for $|\beta|\to\infty$.
	That is, $\tilde\Omega(\beta;w)$ has to suppresses the rising behavior of $\Phi_{\max}(\beta)$ in Eq.~\eqref{eq:MaxIncreaseCF} for any $w>0$ (see also Sec.~\ref{sec:Dual}).

\begin{figure}[ht]
	\includegraphics[width=4.3cm]{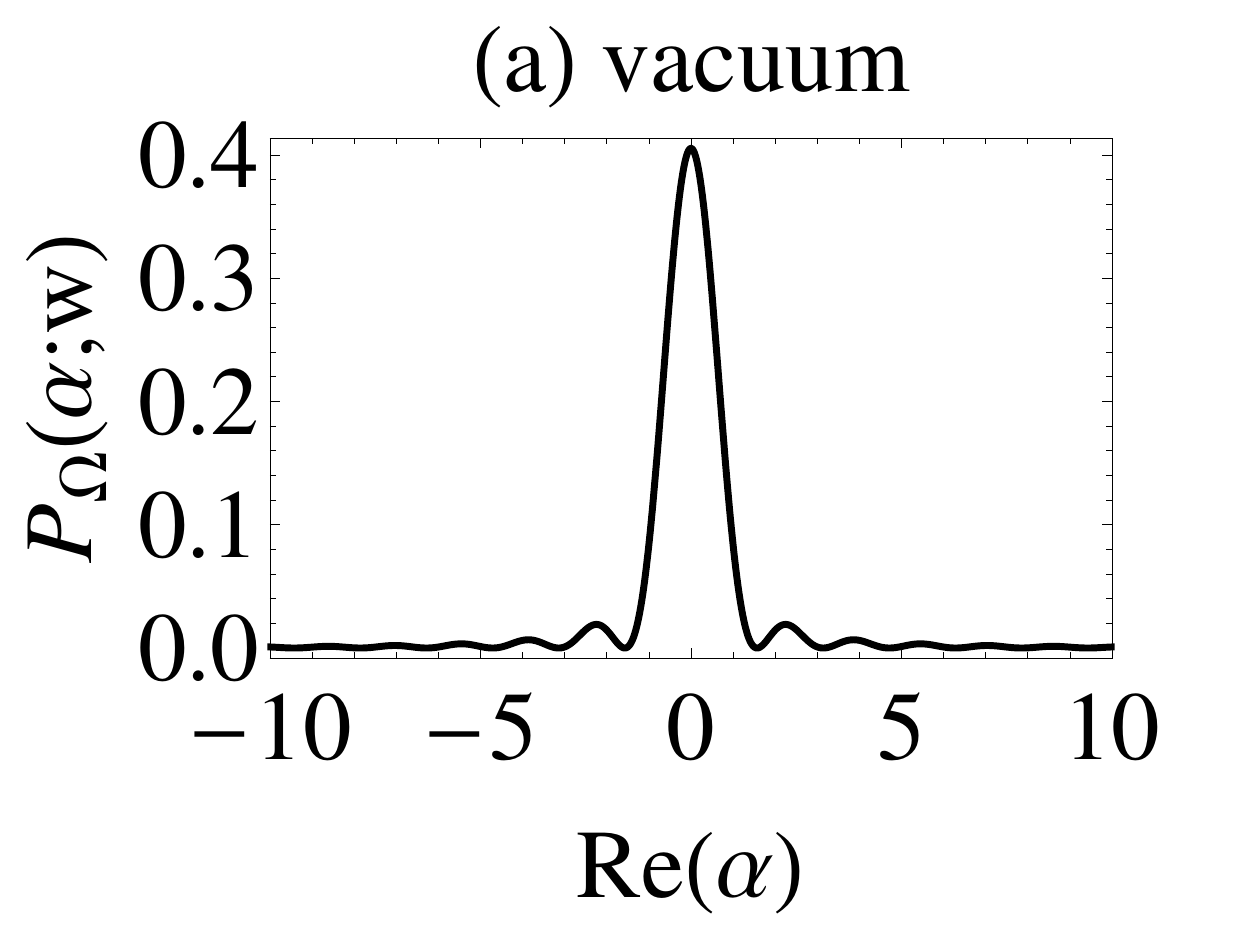}\hfill
	\includegraphics[width=4.3cm]{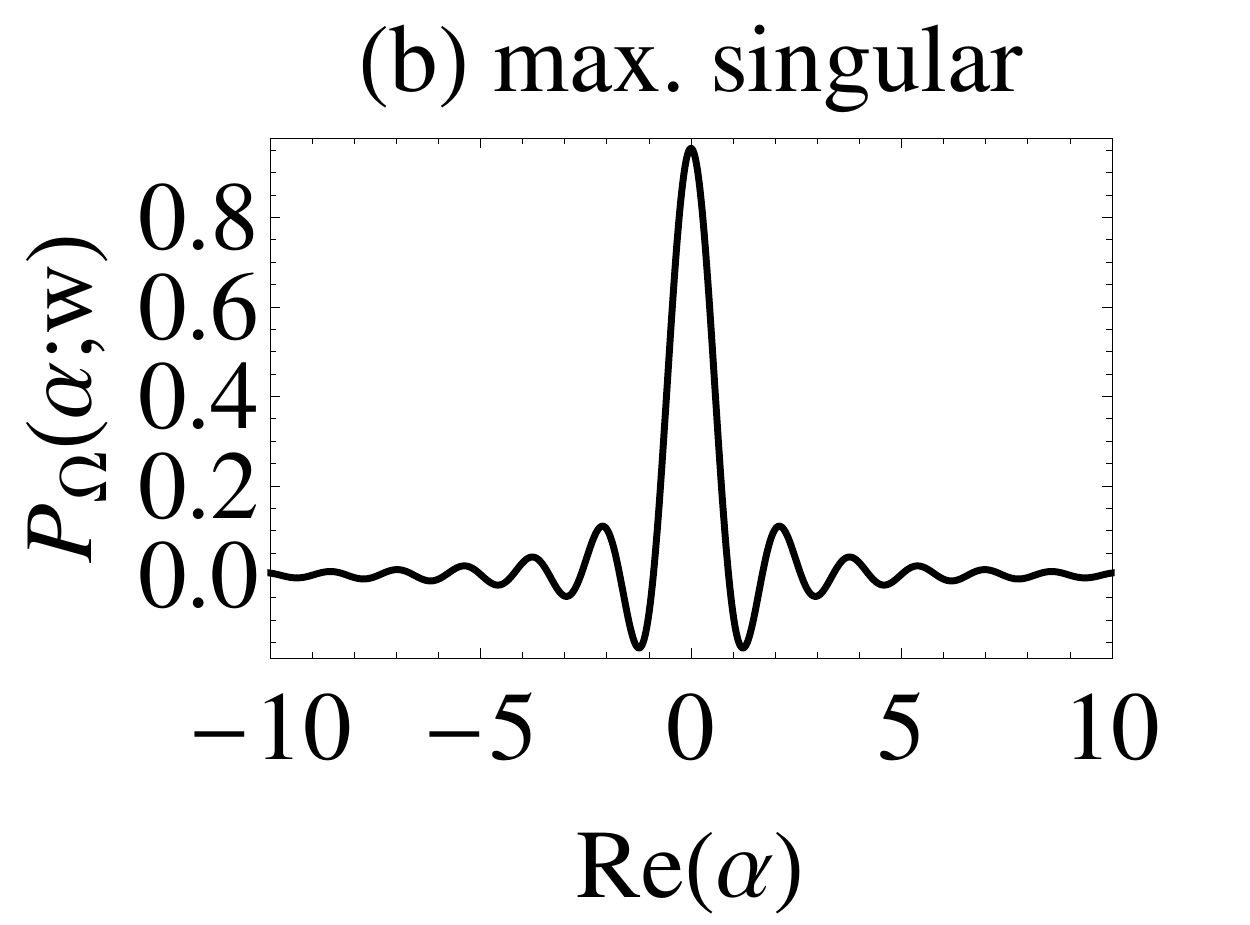}\\[2ex]
	\includegraphics[width=4.3cm]{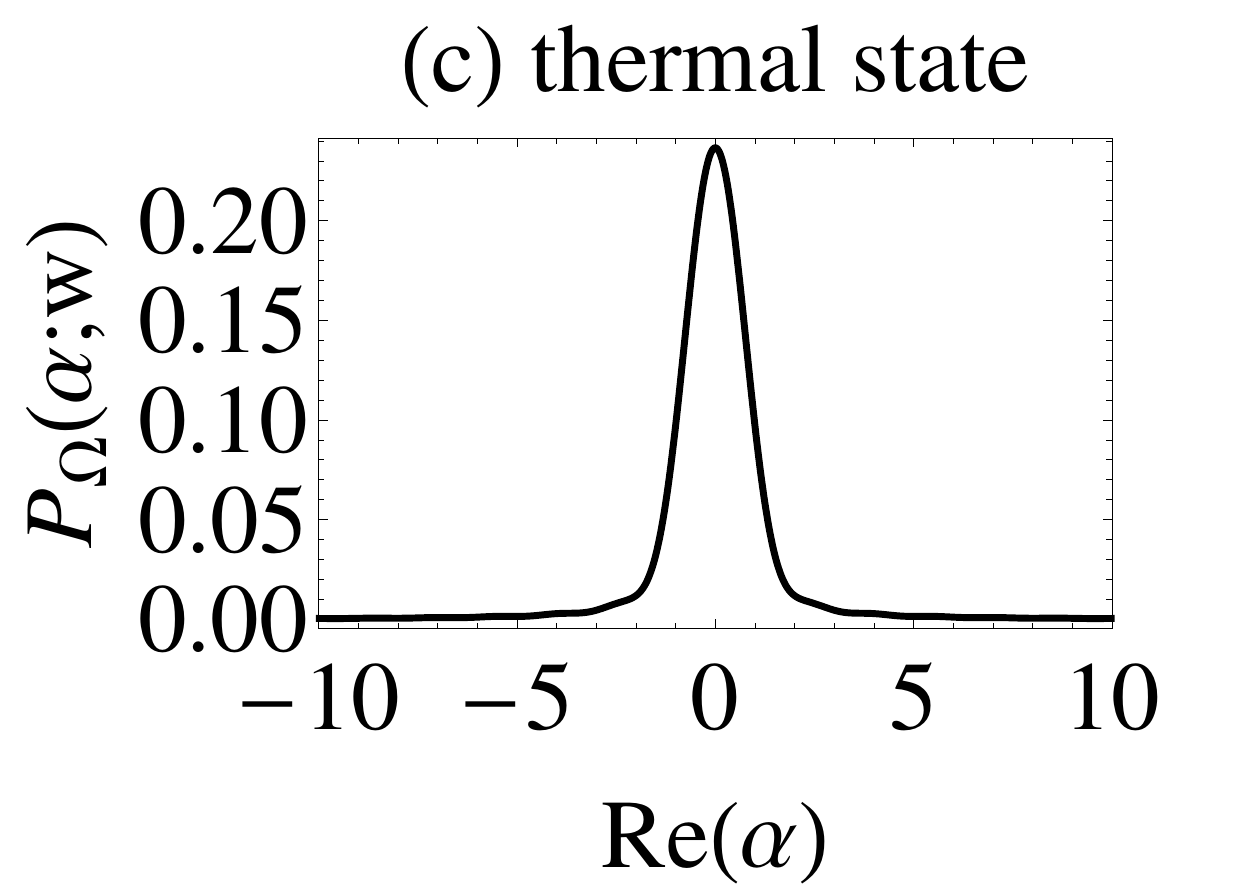}\hfill
	\includegraphics[width=4.3cm]{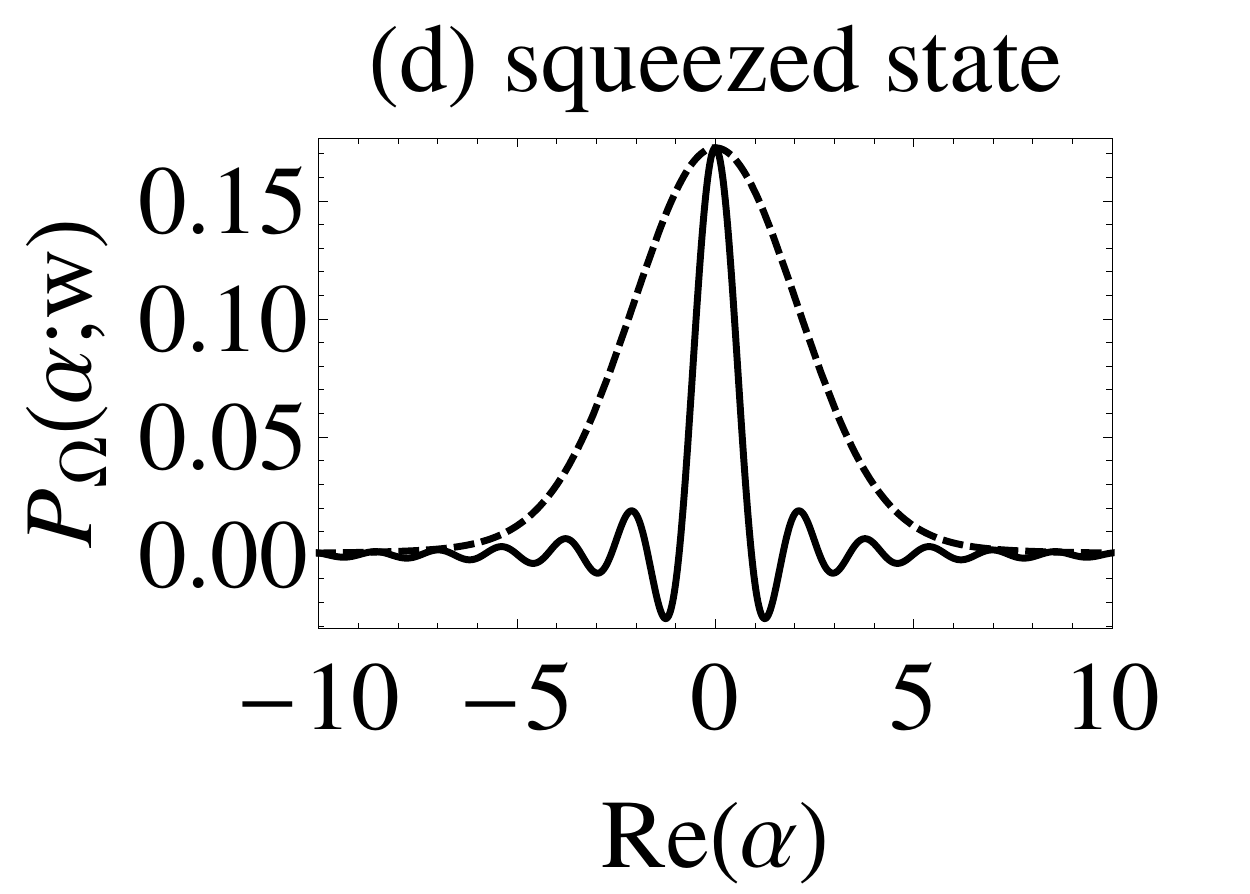}
	\caption{
		Regularized $P_{\Omega}(\alpha;w)$ functions~\eqref{eq:FilterPconv}, for the filter in Eq.~\eqref{sq:SincFilter} with $w=2$, are shown for (a) a vacuum state, (b) the maximally singular GS distribution $P_{\max}$, (c) a thermal state with $\bar n=1/2$, and a squeezed state.
		For the squeezed state, a squeezing parameter $\xi=1.4$ is used and the antisqueezed part is additionally depicted (dashed line).
	}\label{fig:regPexamples}
\end{figure}

	In Fig.~\ref{fig:regPexamples} we show cross sections (${\rm Im}\alpha=0$) of filtered and regular distributions~\eqref{eq:FilterPconv} for different states.
	The applied filter has the form
	\begin{align}\label{sq:SincFilter}
		\Omega(\alpha_0;w)=\frac{w^2}{\pi^2}\left[\frac{\sin(w{\rm Re}\alpha_0)}{w{\rm Re}\alpha_0}\right]^2\left[\frac{\sin(w{\rm Im}\alpha_0)}{w{\rm Im}\alpha_0}\right]^2,
	\end{align}
	which results from a triangular function $\tilde\Omega'$; see Appendix~\ref{app:GaussianFilteredP} for analytical results.
	Figure~\ref{fig:regPexamples}(a) represents a vacuum state as a reference.
	In this case, we have $P(\alpha)=\delta(\alpha)$, which results in $P_{\Omega}(\alpha;w)=\Omega(\alpha;w)\geq0$.

	In Fig.~\ref{fig:regPexamples}(b), we can see the nonclassical behavior of the maximally singular distribution $P_{\max}$.
	Using the filter method, this highly singular distribution transforms into a regular phase-space function with clear negativities.
	In contrast, the thermal state in Fig.~\ref{fig:regPexamples}(c) is described by a classical probability density.
	For this state, a mean thermal photon number $\bar n=1/2$ was chosen as this represents the state in Eq.~\eqref{eq:AbsMu}.
	One can also see that the singularities of the expansion of the thermal state in Eq.~\eqref{eq:ThermalSingular} are correctly converted into a regular and non-negative phase-space function using the regularization~\eqref{eq:FilterPconv}.

	Figure~\ref{fig:regPexamples}(d) shows the squeezed (solid) and antisqueezed (dashed) cuts of the filtered $P_{\Omega}$ function of a squeezed state in Eq.~\eqref{eq:SqStateFock}.
	The highest currently available squeezing level of roughly $12\,{\rm dB}$ ($\xi=1.4$) was taken, as it has been reported in the experiment~\cite{ESBHVMMS10}.
	At this high squeezing level, we can observe that $P_{\Omega}$ for the squeezed part is very similar to the one of the maximally singular phase-space distribution in Fig.~\ref{fig:regPexamples}(b).
	This visualizes our earlier finding that $P_{\max}$ can be approached by squeezed states.
	Additionally note that an experimental reconstruction of $P_{\Omega}$ for a squeezed state was performed in Refs.~\cite{KVHS11,ASVKMH15}.

	It is worth pointing out that the technique of filtered $P_{\Omega}$ functions via non-Gaussian convolution kernels clearly outperforms the $s$ parametrization.
	More precisely, considering the discussion in Sec.~\ref{subsec:sParametrization}, we have seen that the $s$ parametrization of $P_{\max}$ either is a singular distribution ($s>0$) or resembles a regular and non-negative function ($s<0$).
	Using the approach of non-Gaussian filter functions, we can regularize $P_{\max}$ and at the same time identify its nonclassical behavior in terms of negativities.

\section{Nonclassicality tests and test functions}\label{sec:Dual}
	So far, we considered the representation of a quantum state of light, $\hat\rho$, in terms of singular phase-space distributions.
	However, only the expectation value of some observables is experimentally accessible, e.g., in a normally ordered form,
	\begin{align}\label{eq:generalExpectationValue}
		\langle{:}\hat F{:}\rangle={\rm tr}(\hat\rho{:}\hat F{:})=\int d^2\alpha\, P(\alpha)F(\alpha),
	\end{align}
	with $F(\alpha)=\langle\alpha|{:}\hat F{:}|\alpha\rangle$.
	For instance, it is useful to consider measurable nonclassicality criteria to uncover the nonclassicality of a quantum state.
	In particular, one can study the negativity of normally ordered expectation values~\cite{TG65,M86}.
	Such nonclassicality criteria are usually written as
	\begin{align}\label{eq:generalNclCondition}
		0>\langle{:}\hat f^\dag\hat f{:}\rangle,
	\end{align}
	for an arbitrary operator function $\hat f$ depending on $\hat a$ and $\hat a^\dag$.
	This means that we chose ${:}\hat F{:}={:}\hat f^\dag\hat f{:}$.
	An expansion of $\hat f$ in a Taylor series yields the moment-based criteria~\cite{SV05} or a Fourier expansion results in characteristic-function-based criteria~\cite{RV02}.
	Even a combination of both approaches is possible for a unified classification of nonclassicality~\cite{RSAMKHV15}.
	Rewriting~\eqref{eq:generalNclCondition} in terms of the $P$ distribution, we get
	\begin{align}\label{eq:generalNclConditionP}
		0>\langle{:}\hat f^\dag\hat f{:}\rangle=\int d^2\alpha\, P(\alpha)|f(\alpha)|^2,
	\end{align}
	where $f(\alpha)$ is the function that replaces all operators $\hat a$ and $\hat a^\dagger$ in $\hat f$ with $\alpha$ and $\alpha^\ast$, respectively,
	or $F(\alpha)=|f(\alpha)|^2$ in relation to Eq.~\eqref{eq:generalExpectationValue}.
	The resulting nonclassicality criteria have been extensively studied; see, e.g.,~\cite{MBWLN10,VW06}.
	In addition, they have also been related to Hilbert's 17th problem~\cite{KCWL05}.

	As physical measurements should give a finite result, $|\langle{:}\hat F{:}\rangle|<\infty$ in Eq.~\eqref{eq:generalExpectationValue}, we will focus on the possible test functions $F(\alpha)$ from the distributional point of view.
	In order to find such suitable test functions $F(\alpha)$, let us apply our worst-case GS distribution in Eq.~\eqref{eq:Pmax} to Eq.~\eqref{eq:generalExpectationValue}:
	\begin{align}\nonumber
		\langle{:}\hat F{:}\rangle_{\max}=&\int d^2\alpha\, P_{\max}(\alpha)F(\alpha)
		\\=&\sum_{n=0}^\infty\frac{(-1/2)^n}{n!}\left[\partial_\alpha^n\partial_{\alpha^\ast}^n F(\alpha)\right]_{\alpha=0}.
		\label{eq:maxPexpectation}
	\end{align}
	It can be readily seen that the function $F$ has to be a smooth one.
	This means that all derivatives at one point (here for $\alpha=0$ and via displacement for any $\alpha$) have to exist.
	This also relates to the properties of the filter function in Sec.~\ref{sec:FilteredP}.
	That a smooth function $F(\alpha)$ is not sufficient will be studied in this final section along with the formulation of sufficient restrictions to the set of proper test functions.

\subsection{More than analytic test functions}
	In order to apply the nonclassicality of a system via inequality~\eqref{eq:generalNclCondition}, it is therefore indispensable to identify the corresponding class of suitable test functions.
	For this reason, let us consider the upper bound of the expectation value~\eqref{eq:maxPexpectation},
	\begin{align}\label{eq:TestFctBound}
		\left|\langle{:}\hat F{:}\rangle_{\max}\right|\leq \sum_{n=0}^\infty\frac{(1/2)^n}{n!}\left|\partial_\alpha^n\partial_{\alpha^\ast}^n F(\alpha)\right|_{\alpha=0}.
	\end{align}
	Note that this bound is tight, which can be seen for functions with derivatives that have alternating signs, i.e., $\mathrm{sign}[\partial_\alpha^n\partial_{\alpha^\ast}^n F(\alpha)|_{\alpha=0}]=(-1)^n$.
	It has been outlined in many works (e.g., see the discussion presented in~\cite{BV87} and the references therein) that the $P$ distribution can have singularities beyond tempered distributions.
	Thus, a test function in the Schwartz space, i.e., a smooth and rapidly decaying function~\cite{Y80}, is not sufficient.

	Let us consider analytic functions in the next step.
	In contrast to smooth (derivatives of all orders exists), an analytic function is smooth and the Taylor series converges.
	Let us remember that $F(\alpha)$ is a function of the two real parameters ${\rm Re}\alpha$ and ${\rm Im}\alpha$ (see also Appendix~\ref{app:mathQO} in this context).
	Equivalent to the definition of an analytic function, having a locally converging Taylor series, one can also take the criterion in~\cite{K60}.
	Here this criterion applies as follows:
	For the neighborhood of $\alpha=0$ there exist non-negative constants $M$ and $C$ such that the derivatives are bounded as
	\begin{align}\label{eq:analytic}
		\left|\partial_\alpha^m\partial_{\alpha^\ast}^nF(\alpha)\right|\leq M C^{m+n}m!n!,
	\end{align}
	if and only if $F$ is an analytic function.
	Inserting this into inequality~\eqref{eq:TestFctBound}, we get
	\begin{align}
		\left|\langle{:}\hat F{:}\rangle_{\max}\right|\leq
		M\sum_{n=0}^\infty n!\left(\frac{C^2}{2}\right)^n,
	\end{align}
	which diverges for any $C\neq0$, as one can see from Stirling's formula for the asymptotic approximation the factorial, $n!\sim\sqrt{2\pi n}(n/e)^n$.
	Thus, the desired class of physical test functions, $|\langle{:}\hat F{:}\rangle_{\max}|<\infty$, has to be even more regular than analytic functions.

	In order to ensure the convergence in Eq.~\eqref{eq:TestFctBound} and to retrieve well-defined nonclassicality probes, $F(\alpha)=|f(\alpha)|^2$, let us formulate a class of proper test functions.
	Taking the Taylor expansion
	\begin{align}\label{eq:TaylorTest}
		F(\alpha)=\sum_{m,n=0}^\infty \frac{[\partial_\alpha^m\partial_{\alpha^\ast}^nF(\alpha)]_{\alpha=0}}{m!n!}\,\alpha^m\alpha^{\ast n}
	\end{align}
	into account, the criterion~\eqref{eq:analytic} for an analytic function gives
	\begin{align}
		|F(\alpha)|\leq \frac{M}{(1-|\alpha|C)^2}.
	\end{align}
	for $\alpha$ values with $C|\alpha|<1$ and applying $\sum_{n=0}^\infty x^n=(1-x)^{-1}$.
	To bound the right-hand side of~\eqref{eq:TestFctBound}, we may proceed similarly to this idea for analytic functions.
	Namely, we formulate the sufficient constraint that we require the existence of a constant $0\leq C<1$ such that
	\begin{align}\label{eq:strongeranalytical}
		\left|\partial_\alpha^n\partial_{\alpha^\ast}^m F(\alpha)\right|\leq \left(\sqrt 2C\right)^{n+m} \sqrt{n!m!}
	\end{align}
	in the neighborhood of $\alpha=0$ and for all orders of derivatives $m$ and $n$.
	In such a case, we can bound
	\begin{align}
		\left|\langle{:}\hat F{:}\rangle_{\max}\right|\leq\frac{1}{1-C^2}.
	\end{align}

	Moreover, in Appendix~\ref{app:convR} it is shown that the Taylor series~\eqref{eq:TaylorTest} of our considered class of test functions, which satisfy~\eqref{eq:strongeranalytical}, converges for any $\alpha$.
	Thus, the Taylor series describes $F(\alpha)$ globally and not only in the neighborhood of $\alpha=0$.
	This also confirms that the expression in Eq.~\eqref{eq:maxPexpectation} is not only depending on local properties of $F(\alpha)$.
	In addition to the finite derivative scenario in Eq.~\eqref{eq:finiteDual}, the Taylor series discussed  here completely describes the function.
	Again, this underlines that $P_{\max}$ addresses global properties, which is in contrast to distributions with a finite order of derivatives of the Dirac $\delta$~distribution.
	The latter ones are the typically discussed examples in the context of singular GS distributions.
	However, the general complexity of irregular distributions in quantum optics can be studied with our maximally singular GS distribution $P_{\max}$.

	Let us also mention that the constraint~\eqref{eq:strongeranalytical} is stronger (more restrictive) than the constraint~\eqref{eq:analytic}.
	Thus, the family of test functions that fulfill the former conditions are at least analytic and thus smooth functions.
	As final remark, note that the functions $F(\alpha)$, satisfying condition~\eqref{eq:strongeranalytical}, are certainly in the desired class of suited test functions of GS distributions.
	However, there might be other test functions that do not fulfill that requirement, although any expectation value~\eqref{eq:generalExpectationValue} for any GS distribution is finite.
	Thus, proving that condition~\eqref{eq:strongeranalytical} is also necessary requires further investigation.

\subsection{Outlook: Other types of singular behavior}
	In this work, we focused on the singular behavior of the $P$ distribution in terms of derivatives of the Dirac $\delta$~distribution.
	However, one can define other notions of singularities that are connected to other aspects of a distribution that is applied to general test functions.
	Let us consider such an example.

	We may study a family of regular and classical $P$ functions.
	Those are
	\begin{align}\label{eq:LorenzDist}
		P_{\rm cl}(\alpha;t)=\frac{t}{\pi}\frac{1}{(1+|\alpha|^2)^{1+t}},
	\end{align}
	for $t>0$.
	The form of singularity considered here is given by the asymptotic behavior
	\begin{align}
		P_{\rm cl}(\alpha;t)\sim\frac{t}{\pi}\frac{1}{|\alpha|^{2(1+t)}}
	\end{align}
	for large coherent amplitudes $|\alpha|\gg 1$.
	Let us compute the expectation value of test functions $f(\alpha)$ being $n$th-order polynomials.
	It is sufficient to consider monomials $F(\alpha)=|f(\alpha)|^2=|\alpha|^{2n}$ that yield ($r=|\alpha|$ and $R\gg1$)
	\begin{align}\label{eq:LorenzDistMoments}
		&\nonumber \int d^2\alpha\, P_{\rm cl}(\alpha;t)|\alpha|^{2n}=\int_0^\infty dr\, \frac{2t r^{2n+1}}{(1+r^2)^{t+1}}
		\\\sim&\int_{0}^R dr\, \frac{2t r^{2n+1}}{(1+r^2)^{t+1}}+2t\int_R^\infty dr\, \frac{1}{r^{2(t-n)+1}},
	\end{align}
	where the latter asymptotic form takes a finite value if $t>n$.
	For $t\leq n$ the integral diverges.

	A nonclassical state with such a type of asymptotic behavior is
	\begin{align}\label{eq:LorenzDistNcl}
		P_{\rm ncl}(\alpha;t)=\frac{P_{\rm cl}(\alpha;t)-\mathcal N_t\delta(\alpha)}{1-\mathcal N_t},
	\end{align}
	with $\mathcal N_t=\int d^2\alpha\, P_{\rm cl}(\alpha;t)\exp(-|\alpha|^2)$.
	The state is constructed in analogy to the discussion in Ref.~\cite{DV00}.
	The corresponding density operator $\hat\rho$ is diagonal in the Fock basis [$P_{\rm ncl}(\alpha;t)$ does not depend on the phase $\arg\alpha$] and it has no vacuum contribution, $\langle{\rm vac}|\hat\rho|{\rm vac}\rangle=0$.
	As for any classical state $\langle{\rm vac}|\hat\rho|{\rm vac}\rangle=\int d^2\alpha\, P(\alpha)\exp(-|\alpha|^2)>0$ holds, this is a nonclassical state~\cite{DV00}.

	Let us briefly discuss these results.
	The given family of states~\eqref{eq:LorenzDist} and~\eqref{eq:LorenzDistNcl} cannot be completely characterized by their moments, as most of them do not exist.
	Due to their asymptotic behavior, these functions could also be considered as singular distributions.
	In particular, the example $t=1$ is interesting from the physical point of view.
	Quantum states that are described by this classical Cauchy-Lorentz distribution, $t=1$ in Eq.~\eqref{eq:LorenzDist}, have an infinite energy, because of an infinite photon number [$\langle \hat a^\dagger\hat a\rangle=\infty$, i.e., $n=t=1$ in Eq.~\eqref{eq:LorenzDistMoments}].
	Despite the fact that the $P$ function is a regular and classical one, they cannot be generated by a realistic physical process with a finite energy.
	Moreover, the nonclassicality of the state in Eq.~\eqref{eq:LorenzDistNcl} for $t=1$ cannot be confirmed with the matrix of moments nonclassicality criteria~\cite{SV05}.
	Instead, its nonclassicality can be inferred, for example, with the nonclassicality tests in Ref.~\cite{RV02}.
	This means that the asymptotic behavior and the requirement of physical processes for the state preparation gives another form of singular behavior and restricts the set of possible distributions, respectively.

\section{Summary and Conclusions}\label{sec:SnC}
	In summary, a characterization of the singularities of phase-space distributions has been performed with a special emphasize on the maximally singular Glauber-Sudarshan distribution.
	Different regular and irregular phase-space representations beyond the Glauber-Sudarshan distribution have been studied.
	A sufficient criterion for the dual space, i.e., the family of sufficiently smooth test functions for identifying nonclassicality, has been presented for this physically relevant class of distributions.
	This treatment is relevant from the fundamental point of view as singular distributions describe many physically relevant scenarios in quantum optics.
	From the experimental point of view, measurements with experimentally generated quantum states are typically described by singular Glauber-Sudarshan distributions, e.g., squeezed or Fock states.

	In this work, the general expansion of any quantum state of light was considered in terms of distributions that contain all orders of derivatives of the Dirac $\delta$~distribution.
	The difference between a finite order of derivatives and an infinite order was elaborated.
	Namely, it was shown that a finite order necessarily yields a nonclassical state, whereas an infinite order does not allow such a conclusion, which was further supported with the example of a thermal state.
	Hence, a statement that a highly singular Glauber-Sudarshan distribution describes a nonclassical state is not always true.

	A maximally singular Glauber-Sudarshan distribution was established and its properties have been studied.
	Along with other aspects, it was demonstrated that such a distribution does not describe a quantum state of light.
	However, it can be approached with, for example, squeezed states for any fixed direction in phase space.
	Thus, the maximally singular distribution can be formally considered as a ``state'' with an infinite squeezing in any phase-space direction, which violates the uncertainty principle.
	We also showed that methods that can regularize the singularities of this maximally singular distribution are automatically applicable to any physical quantum state of light.
	Therefore, this single maximally singular distribution serves as a benchmark for any physical state.

	Moreover, a suitable class of test functions (dual space for the considered class of distributions) was derived.
	It has been shown that such test functions have to be even more regular than typically assumed to overcome the singularities of the Glauber-Sudarshan distribution.
	Finally, other types of irregularities have been discussed.

	Let us formulate some additional conclusions.
	The Glauber-Sudarshan representation is a key notion in quantum optics for discerning classical from nonclassical light.
	Thus, it is required to have a profound knowledge of the irregularities that might occur in this phase-space distribution, e.g., for performing an experimental quantum state reconstruction.
	In particular, the given simple examples and counter-examples underline the fact that an intuition of singularities might lead to incomplete or even incorrect interpretations of the quantum nature of the system under study.
	Thus, the present work might lead to a deeper understanding of the nonclassical effects in quantized radiation fields.

%%%%%%%%%%%%%%%%%%%%%%%%%%%%%%%%%%%%%%%%%%%%%%%%%%%%%%%%%%%%%%%%%%%%%%%%%%%%%%%%%%%%%%%%%%%%%%%%%%%%%%%%%%%%%%%%%%%%%%%%%%%%%%%%
%%%%%%%%%%%%%%%%%%%%%%%%%%%%%%%%%%%%%%%%%%%%%%%%%%%%%%%%%%%%%%%%%%%%%%%%%%%%%%%%%%%%%%%%%%%%%%%%%%%%%%%%%%%%%%%%%%%%%%%%%%%%%%%%
\subsection*{Acknowledgements}
	This work has received funding from the European Union's Horizon 2020 research and innovation program under grant agreement No 665148.
	This work is dedicated to Professor Werner Vogel, Universit\"at Rostock, who is contributing to the field of theoretical quantum optics for more than 40 years.
	I want to thank Werner for all the enlightening discussions and for supporting me.
	I also acknowledge helpful comments by E. Agudelo.

%%%%%%%%%%%%%%%%%%%%%%%%%%%%%%%%%%%%%%%%%%%%%%%%%%%%%%%%%%%%%%%%%%%%%%%%%%%%%%%%%%%%%%%%%%%%%%%%%%%%%%%%%%%%%%%%%%%%%%%%%%%%%%%%
%%%%%%%%%%%%%%%%%%%%%%%%%%%%%%%%%%%%%%%%%%%%%%%%%%%%%%%%%%%%%%%%%%%%%%%%%%%%%%%%%%%%%%%%%%%%%%%%%%%%%%%%%%%%%%%%%%%%%%%%%%%%%%%%
\appendix

%%%%%%%%%%%%%%%%%%%%%%%%%%%%%%%%%%%%%%%%%%%%%%%%%%%%%%%%%%%%%%%%%%%%%%%%%%%%%%%%%%%%%%%%%%%%%%%%%%%%%%%%%%%%%%%%%%%%%%%%%%%%%%%%
%%%%%%%%%%%%%%%%%%%%%%%%%%%%%%%%%%%%%%%%%%%%%%%%%%%%%%%%%%%%%%%%%%%%%%%%%%%%%%%%%%%%%%%%%%%%%%%%%%%%%%%%%%%%%%%%%%%%%%%%%%%%%%%%
\section{Calculus with complex amplitudes}\label{app:mathQO}
	In quantum optics, we have a particular mathematical analysis with coherent amplitudes $\alpha$, where $\alpha$ is a complex number that may be decomposed into its real part $x=(\alpha+\alpha^\ast)/2$ and imaginary part $p=(\alpha-\alpha^\ast)/(2i)$.
	In order to avoid confusion with complex analysis and for our rigorous analysis, let us recall some known but important aspects of this treatment.
	For example, the function $|\alpha|^2$ is not differentiable in complex analysis, but in quantum optics we have: $\partial_\alpha|\alpha|^2=\alpha^\ast$ or $\partial_{\alpha^\ast}|\alpha|^2=\alpha$.

	We may identify a quantum optical function $f(\alpha)$ with a function of two real-valued variables
	\begin{align}\label{eq:ComplexIdent}
		f(\alpha)=f_2(x,p).
	\end{align}
	The Dirac $\delta$~distribution takes the form
	\begin{align}
		\delta(\alpha-\alpha_0)=\delta(x-x_0)\delta(p-p_0),
	\end{align}
	with $\alpha_0=x_0+ip_0$.
	The derivatives can be written correspondingly as
	\begin{align}\label{eq:ComplexPartial}
		\partial_\alpha=\frac{1}{2}\left(\partial_x-i\partial_p\right)
		\text{ and }\partial_{\alpha^\ast}=\frac{1}{2}\left(\partial_x+i\partial_p\right),
	\end{align}
	which follows from Eq.~\eqref{eq:ComplexIdent} or the identification $f(\alpha)=f_2([\alpha+\alpha^\ast]/2,-i[\alpha-\alpha^\ast]/2)$.
	Furthermore, the Fourier transform is
	\begin{align}
	\begin{aligned}
		&\int d^2\alpha\, f(\alpha)e^{\beta\alpha^\ast-\beta^\ast\alpha}
		\\=&\int dxdp\, f_2(x,p) e^{i(2{\rm Im}\beta)x}e^{-i(2{\rm Re}\beta)p},
	\end{aligned}
		\label{eq:ComplexFT}
	\end{align}
	where we used $d^2\alpha=dx\,dp$.
	In the following we will apply this formulation in terms of two real values rather than the complex coherent amplitude.
	In this sense one can see that, e.g., $\partial_\alpha|\alpha|^2=\alpha^\ast$ ($\partial_{\alpha^\ast}|\alpha|^2=\alpha$) is properly retrieved.

%%%%%%%%%%%%%%%%%%%%%%%%%%%%%%%%%%%%%%%%%%%%%%%%%%%%%%%%%%%%%%%%%%%%%%%%%%%%%%%%%%%%%%%%%%%%%%%%%%%%%%%%%%%%%%%%%%%%%%%%%%%%%%%%
%%%%%%%%%%%%%%%%%%%%%%%%%%%%%%%%%%%%%%%%%%%%%%%%%%%%%%%%%%%%%%%%%%%%%%%%%%%%%%%%%%%%%%%%%%%%%%%%%%%%%%%%%%%%%%%%%%%%%%%%%%%%%%%%
\section{Exponential Laplace operator}\label{app:ExponentialLaplace}
	Throughout this work, we encountered derivatives of the form $\partial_\alpha\partial_{\alpha^\ast}$.
	Such an operator can be identified with the two-dimensional Laplace operator $\Delta_{x,p}=\partial_x^2+\partial_p^2$, as we have
	\begin{align}
		\partial_\alpha\partial_{\alpha^\ast}f(\alpha)=\frac{1}{4}\Delta_{x,p}f_2(x,p),
	\end{align}
	where we applied Eq.~\eqref{eq:ComplexPartial}.
	The operator $-\Delta_{x,p}$ is a positive-semi-definite operator.
	From the bounded eigenfunctions, we get the non-negative eigenvalues
	\begin{align}
		-\Delta_{x,p}e^{ik_xx-ik_pp}=\left(k_x^2+k_p^2\right)e^{ik_xx-ik_pp},
	\end{align}
	with real values $k_x$ and $k_p$ [cf. also Eq.~\eqref{eq:ComplexFT}].
	This means that the exponential Laplace operator
	\begin{align}\label{eq:expLaplace}
		e^{-4t\partial_\alpha\partial_{\alpha^\ast}}=e^{-t\Delta_{x,p}}
	\end{align}
	has for $t>0$ ($t<0$) the eigenvalues $\exp(t[k_x^2+k_p^2])$, which are greater (less) than one.
	For $t=0$, one can set $\exp(-t\Delta_{x,p})=1$.
	This means that the exponential Laplace operator~\eqref{eq:expLaplace} describes the identity for $t=0$, a contractive map for $t<0$, and an expansive map for $t>0$.
	Note that this relates to the classical or nonclassical features of $\exp(-4t\partial_\alpha\partial_{\alpha^\ast})\delta(\alpha)$ for different $t$ values.

	Further, let us formulate the $P$ distribution of the pure squeezed state with the characteristic function in Eq.~\eqref{eq:SqueezedStateCF}.
	The identification $\beta=k_x+ik_p$ allows one to compute
	\begin{align}
		P(\alpha)=&\frac{1}{\pi^2}\int dk_xdk_p \exp\left(i[2{\rm Im}\alpha]k_x-i[2{\rm Re}\alpha]k_p\right)\nonumber
		\\&\times\exp\left(-\frac{e^{2\xi}-1}{2}k_x^2+\frac{1-e^{-2\xi}}{2}k_p^2\right)\nonumber
		\\=&\sqrt{\frac{2}{\pi\left(e^{2\xi}-1\right)}}\exp\left(-\frac{2[{\rm Im}\alpha]^2}{e^{2\xi}-1}\right)\label{eq:antisqP}
		\\&\times\exp\left(-\frac{1-e^{-2\xi}}{2}\frac{\partial_{{\rm Re}\alpha}^2}{4}\right)\delta({\rm Re}\alpha)\label{eq:sqP}.
	\end{align}
	In this form, the third line~\eqref{eq:antisqP} represents the antisqueezed direction ${\rm Im}\alpha$, which is described by a regular Gaussian distribution,
	and the fourth line~\eqref{eq:sqP} corresponds to the squeezed direction ${\rm Re}\alpha$, which resembles an exponential second derivative of the one-dimensional Dirac $\delta$~distribution.

%%%%%%%%%%%%%%%%%%%%%%%%%%%%%%%%%%%%%%%%%%%%%%%%%%%%%%%%%%%%%%%%%%%%%%%%%%%%%%%%%%%%%%%%%%%%%%%%%%%%%%%%%%%%%%%%%%%%%%%%%%%%%%%%
%%%%%%%%%%%%%%%%%%%%%%%%%%%%%%%%%%%%%%%%%%%%%%%%%%%%%%%%%%%%%%%%%%%%%%%%%%%%%%%%%%%%%%%%%%%%%%%%%%%%%%%%%%%%%%%%%%%%%%%%%%%%%%%%
\section{Analytic filtered $P$ function for Gaussian states}\label{app:GaussianFilteredP}
	For the considered considered in Fig~\ref{fig:regPexamples}, let us derive analytic expressions for the filtered and regular phase-space representation $P_{\Omega}(\alpha;w)$.
	In a compact form, the definition reads~\cite{KV10}
	\begin{align}\label{Eq:FilterCF1}
	\begin{aligned}
		P_\Omega(\alpha;w)=&\frac{1}{\mathcal N}\int d^2\beta\, e^{\beta^\ast\alpha-\beta\alpha^\ast}\Phi(\beta)
		\\&\phantom{\frac{1}{\mathcal N}}\times
		\int d^2\beta_0\,\omega(\beta/w+\beta_0)\omega(\beta_0)^\ast,
		\\\text{with }
		\mathcal N=&\pi^2\int d^2\beta_0\,|\omega(\beta_0)|^2.
	\end{aligned}
	\end{align}
	Let us take the first example of a function $\omega$ that was proposed in Ref.~\cite{KV10}: $\omega(\beta)=1$ for $-1/2\leq{\rm Re}\beta\leq 1/2$ and $-1/2\leq{\rm Im}\beta\leq 1/2$, and otherwise $\omega(\beta)=0$.
	This yields $\int d^2\beta_0\,\omega(\beta+\beta_0)\omega(\beta_0)^\ast={\rm tri}({\rm Re}\beta){\rm tri}({\rm Im}\beta)$, where the triangular function is given by
	\begin{align}
		{\rm tri}(x)=\left\lbrace\begin{array}{ll}
			0 &\text{for $x\leq-1$,}\\
			1+x &\text{for $-1\leq x\leq 0$,}\\
			1-x &\text{for $0\leq x\leq 1$,}\\
			0 &\text{for $1\leq x$.}\\
		\end{array}\right.
	\end{align}
	IN particular, this is a real-valued function with ${\rm tri}(0)=1$ and ${\rm tri}(-x)={\rm tri}(x)$.
	Moreover, the examples of Gaussian characteristic functions can be written as
	\begin{align}
		\Phi(\beta)=\exp\left(-\lambda x^2-\kappa p^2\right),
	\end{align}
	with $x={\rm Re}\beta$, $p={\rm Im}\beta$, and real numbers $\kappa$ and $\lambda$.
	Note that a displacement and rotations in phase space could also be included.
	Now, Eq.~\eqref{Eq:FilterCF1} reduces to
	\begin{align}\label{Eq:FilterCF2}
	\begin{aligned}
		P_{\Omega}(\alpha;w)=&\int\frac{dx}{\pi}\,e^{
			2i({\rm Im}\alpha)x
			-\lambda x^2
		}{\rm tri}\left(\frac{x}{w}\right)
		\\&\times
		\int\frac{dp}{\pi}\,e^{
			2i(-{\rm Re}\alpha)p
			-\kappa p^2
		}{\rm tri}\left(\frac{p}{w}\right).
	\end{aligned}
	\end{align}

	We may evaluate the following function:
	\begin{align}\label{eq:auxFct}
		&T(y;g)=\int\frac{dz}{\pi}\,e^{2iyz-gz^2}{\rm tri}(z)
		\\\nonumber=&{\rm Re}\left[\frac{2}{\pi}\int_{0}^1dz\, e^{-gz^2+2iyz}(1-z)\right]
		\\\nonumber=&{\rm Re}\left[\frac{e^{-g+2 i y}-1}{\pi  g}\right.
		\\\nonumber&+\left.\frac{e^{-y^2/g}}{\sqrt{\pi } g}\frac{g-i y}{\sqrt g}
			\left({\rm erf}\left[\frac{g-i y}{\sqrt{g}}\right] - {\rm erf}\left[-\frac{iy}{\sqrt{g}}\right]\right)
		\right]
	\end{align}
	for real values $y$ and $g$.
	The error function is defined as
	\begin{align}
		{\rm erf}(x)&
		=\frac{2}{\sqrt\pi}\int\limits_{0}^xdt\,e^{-t^2}
		=\frac{2}{\sqrt\pi}\sum_{n=0}^\infty\frac{(-1)^nx^{2n+1}}{n!(2n+1)}
	\end{align}
	and it has the properties
	\begin{align}
		{\rm erf}(x)&
		=-{\rm erf}(-x)={\rm erf}(x^\ast)^\ast
	\end{align}
	for every complex number $x$.
	For the specific choice $g=0$, we have
	\begin{align}
		T(y;0)=\frac{1}{\pi}\left[\frac{\sin(y)}{y}\right]^2.
	\end{align}
	Further, the argument $y=0$ yields
	\begin{align}
		T(0;g)=&{\rm Re}\left[
			\frac{e^{-g}-1}{\pi  g}
			+\frac{{\rm erf}(\sqrt g)}{\sqrt{\pi g}}
		\right].
	\end{align}
	Finally, we can express Eq.~\eqref{Eq:FilterCF2} in an analytic form
	\begin{align}
		P_{\Omega}(\alpha;w)=w^2T(w{\rm Im}\alpha;w^2\lambda)T(-w{\rm Re}\alpha;w^2\kappa).
	\end{align}

%%%%%%%%%%%%%%%%%%%%%%%%%%%%%%%%%%%%%%%%%%%%%%%%%%%%%%%%%%%%%%%%%%%%%%%%%%%%%%%%%%%%%%%%%%%%%%%%%%%%%%%%%%%%%%%%%%%%%%%%%%%%%%%%
%%%%%%%%%%%%%%%%%%%%%%%%%%%%%%%%%%%%%%%%%%%%%%%%%%%%%%%%%%%%%%%%%%%%%%%%%%%%%%%%%%%%%%%%%%%%%%%%%%%%%%%%%%%%%%%%%%%%%%%%%%%%%%%%
\section{Convergence radius of test functions}\label{app:convR}
	The Taylor series~\eqref{eq:TaylorTest} of a function that satisfies~\eqref{eq:strongeranalytical} has a radius of convergence $R=\infty$.
	To show this, we may compute
	\begin{align}
		|F(\alpha)|\leq&\sum_{m,n=0}^\infty \frac{(\sqrt 2 C)^{m+n}}{\sqrt{m!n!}}|\alpha|^{m+n}=\sum_{k=0}^\infty c_k |\alpha|^k,
		\\\text{with }c_k&=\sum_{n=0}^k \frac{(\sqrt 2 C)^{k}}{\sqrt{(k-n)!n!}},
	\end{align}
	where we substituted $m+n=k$.
	Note that $c_k\geq0$.
	To get the radius of convergence $R\geq0$, one can apply the root test by Cauchy
	\begin{align}\nonumber
		\frac{1}{R}=&\lim_{k\to\infty}\sup_{l\geq k} c_l^{1/l}
		\\=&\sqrt 2 C\lim_{k\to\infty}\sup_{l\geq k}\left[
			\sum_{n=0}^l \frac{1}{\sqrt{(l-n)!n!}}
		\right]^{1/l}.
	\end{align}
	The inner sum can be interpreted as a 1-norm of an $l$-dimensional vector $x$, which is bounded by the 2-norm via $\|x\|_1\leq \sqrt l \|x\|_2$.
	Thus, using $2^l=\sum_{n=0}^l l!/(n!(l-n)!)$, we can write the upper bound
	\begin{align}\nonumber
		\frac{1}{R}\leq& \sqrt 2 C\lim_{k\to\infty}\sup_{l\geq k}\left[
			\sqrt{\frac{l2^l}{l!}}
		\right]^{1/l}
		\\&=2C\lim_{k\to\infty}\sup_{l\geq k}\left[\frac{1}{(l-1)!}\right]^{1/(2l)}
		=0,
	\end{align}
	where $[(l-1)!]^{-1/(2l)}$ is a monotonically decreasing sequence with the limit 0.
	Therefore, we have $1/R\leq 0$, which means that the radius of convergence is $R=\infty$.

%%%%%%%%%%%%%%%%%%%%%%%%%%%%%%%%%%%%%%%%%%%%%%%%%%%%%%%%%%%%%%%%%%%%%%%%%%%%%%%%%%%%%%%%%%%%%%%%%%%%%%%%%%%%%%%%%%%%%%%%%%%%%%%%
%%%%%%%%%%%%%%%%%%%%%%%%%%%%%%%%%%%%%%%%%%%%%%%%%%%%%%%%%%%%%%%%%%%%%%%%%%%%%%%%%%%%%%%%%%%%%%%%%%%%%%%%%%%%%%%%%%%%%%%%%%%%%%%%

\end{document}